\documentclass{epl}
\title{Universal behavior of spin-mediated energy transport in   
$S=1/2$ chain cuprates: BaCu$_2$Si$_2$O$_7$ as an example}
\shorttitle{Energy transport in $S=1/2$ chain cuprates}

\author{A. V. Sologubenko\inst{1} 
    \and H. R. Ott\inst{1}
    \and G. Dhalenne\inst{2}
    \and A. Revcolevschi\inst{2}
}
\shortauthor{A. V. Sologubenko \etal}

\institute{
  \inst{1} Laboratorium f\"ur Festk\"orperphysik, ETH H\"onggerberg,
CH-8093 Z\"urich, Switzerland\\
  \inst{2} Laboratoire de Physico-Chimie de l'Etat Solide, Universit\'e Paris-Sud,
91405 Orsay, France
}

\pacs{66.70.+f}{Nonelectronic thermal conduction and heat-pulse propagation in solids; thermal
                                         waves}
\pacs{75.40.Gb}{Dynamic properties (dynamic susceptibility, spin waves, spin diffusion, dynamic scaling, etc.)}
\pacs{75.10.Pq}{Spin chain models}

\begin{document}

\maketitle

\begin{abstract}
The thermal conductivity $\kappa$ of the spin-1/2 chain cuprate 
BaCu$_2$Si$_2$O$_7$ was measured along different crystallographic 
directions in the temperature region between 0.5 
and 300~K.  The thermal conductivity along the chain direction
considerably exceeds that along perpendicular directions. Near the 
antiferromagnetic transition at $T_N = 9.2\un{K}$ the data indicates 
enhanced scattering of phonons by critical fluctuations in the spin system.
A comparison of the data above $T_N$ with available results on 
similar materials reveals similarities in 
the main  features of the temperature dependence of the 
mean free path of itinerant spin excitations. 
This universal behavior is most likely caused by the spin-lattice 
interaction.
\end{abstract}

According to theoretical predictions, the energy transport in several 
types of $S=1/2$ one-dimensional (1D) antiferromagnetic (AFM) spin systems 
is expected to be remarkably different from 
similar three-dimensional (3D) systems. 
In 3D systems, the thermal conductivity of magnetic origin can be large only much below the 
long-range ordering 
transitions, where the itinerant spin excitations are  well-defined quasiparticles (magnons). 
At temperatures close to but below 
the 3D ordering transition, the spin-mediated thermal 
conductivity $\kappa_{s}$ is considerably reduced because of enhanced 
magnon-magnon scattering \cite{Kawasaki63,Stern65}. 
Above the transition, 
the energy transport by spin excitations is small and has diffusive 
character, such that in real materials other carriers of energy, most 
notably phonons and, in electrical conductors, electronic quasiparticles, 
dominate the heat transport. In strictly 1D spin 
systems, no long-range order exists at nonzero temperatures, and 
naively one 
could  expect a rather small thermal conduction via the spin system.
However, for 1D $S=1/2$ spin systems described by the Heisenberg AFM $XXZ$  
Hamiltonian
\begin{equation}\label{Hamiltonian}
H = J \sum_{i} ( S^{x}_{i}S^{x}_{i+1} + S^{y}_{i}S^{y}_{i+1} + 
\Delta S^{z}_{i}S^{z}_{i+1})  
\end{equation}
with $-1 < \Delta \leq 1$, the heat transport turns out to be of ballistic type 
and therefore, $\kappa_{s}$ is infinite for infinite-length 
systems.  Anomalous behavior of energy transport in Heisenberg $S=1/2$ 
AFM XXZ chains has been predicted a long time ago
\cite{Niemeijer71,Krueger71}. It was recently associated with
the integrability of the underlying model, because the energy density appears 
to be one of  the numerous constants of motion of integrable 
systems \cite{Zotos97}.
Very recently, thermal transport in integrable models of several 1D spin 
systems was analyzed employing the Kubo formalism and 
adopting the notion of a ``thermal'' Drude weight with a similar
meaning as the Drude weight in the standard theory of electrical 
transport \cite{Kluemper02,Alvarez02_Ano}. The thermal conductivity 
can thus  be written as  \cite{Alvarez02_Ano}
\begin{equation}\label{KsDrude}
 \kappa_{s}(T) = \kappa^{th}(T) \tau,
\end{equation}
where $\kappa^{th}$ is the thermal Drude weight 
(in ref.~\cite{Kluemper02}  the thermal Drude weight is denoted as 
$\tilde{\kappa} \equiv \pi \kappa^{th}$) 
and $\tau$ is the 
lifetime of the eigenstates of the spin system.  
The Drude-weight can be calculated exactly, e.g., by the Bethe Ansatz 
method. For a system described by the Hamiltonian in eq.~(\ref{Hamiltonian}), 
different methods of calculation give consistent values of $\kappa^{th}(T)$ 
\cite{Kluemper02,Alvarez02_Low,Alvarez02_Ano,Alvarez02_Con,Saito02,HeidrichMeisner02}.
The relaxation time $\tau$ is infinite for an ideal integrable 1D system 
but turns finite 
if the influence of external perturbations, such as defects, phonons, interchain 
coupling, cannot be neglected. 
Theoretical work concerning 
the features of $\tau$ for different types of perturbation is scarce. 
Below, we describe an experimental attempt to investigate details 
of energy transport relaxation in 1D Heisenberg $S=1/2$ $XXZ$ 
model materials.

We have measured the 
thermal conductivity $\kappa$ of the $S=1/2$ spin-chain compound BaCu$_2$Si$_2$O$_7$ 
along different crystallographic directions and at temperatures 
between 0.5 and 300~K. 
BaCu$_2$Si$_2$O$_7$ has an orthorhombic crystal structure with lattice 
parameters $a=6.862\un{\AA}$, $b=13.178\un{\AA}$, 
and  $c=6.897\un{\AA}$ \cite{Oliveira93}. Important structural elements are 
chains of  corner-sharing CuO$_4$ tetrahedra, running along the 
$c$-axis.  Cu$^{2+}$ ions with spins $S=1/2$ are connected along the 
chain direction via 124$^{\circ}$ Cu--O--Cu bonds.
The magnetic properties of individual chains can be well described by the 
isotropic form of the Hamiltonian 
of eq.~(\ref{Hamiltonian}), i.e.,  $\Delta=1$ and $J = 24.1\un{meV}$ 
\cite{Tsukada99}. The interchain interactions are two orders of 
magnitude smaller \cite{Kenzelmann01} and lead to a 3D AFM ordering 
transition at $T_{N} = 9.2\un{K}$. Below $T_{N}$, 
the spins are oriented along the $c$-axis. 
Applying a magnetic field along the 
easy axis leads to an exotic two-stage spin-flop transition  
\cite{Tsukada01}; the origin of this anomalous behaviour is not completely 
understood\cite{Zheludev01_Str,Poirier02}. 
Recent interest in BaCu$_2$Si$_2$O$_7$ was also stimulated 
by the speculation that below $T_{N}$ the ``longitudinal mode'', a 
long-lived spin wave with a polarization parallel to the direction 
of the ordered moments, might be observed \cite{Zheludev02_Dom}.

Three bar-shaped samples with typical dimensions of about 
$0.4 \times 0.5 \times 2\un{mm^{3}}$ were cut 
from the same single crystal grown using the floating-zone image-furnace 
technique. For each sample, the longest dimension was oriented along one 
of the main crystallographic axes. The thermal conductivity was measured 
employing the standard uniaxial heat flow technique. The heat input 
into the 
sample was provided by the Joule heat of a RuO$_{2}$ resistor. The temperature 
gradient was monitored using a system of Chromel-Au+0.07\%~Fe 
thermocouples in the temperature region between 2 and 300~K, and by a 
couple of RuO$_{2}$ thermometers in the temperature regime between 0.5 
and 3~K.

\begin{figure}
\onefigure[width=0.75\columnwidth]{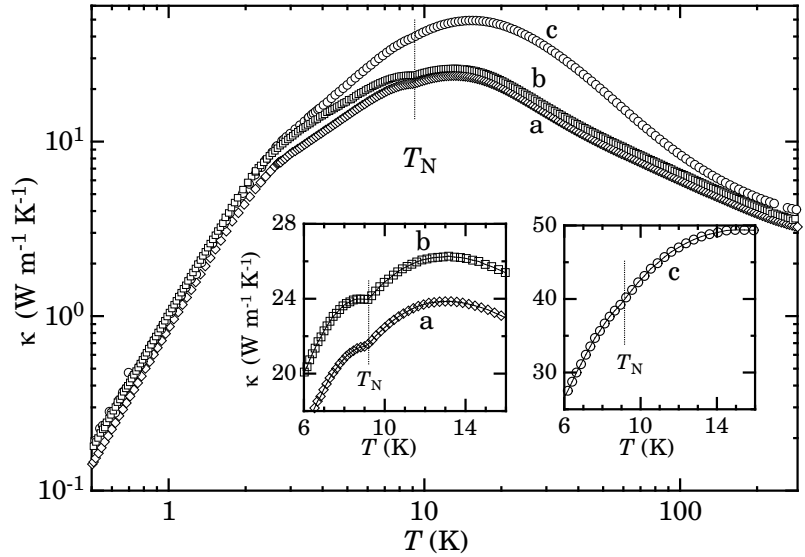}
\caption{Temperature dependences of the thermal conductivities of 
BaCu$_2$Si$_2$O$_7$ along the $a$, $b$, and $c$ axes. The insets emphasize 
the $\kappa(T)$ features near the 3D ordering transition $T_{N} = 
9.2\un{K}$. }
\label{K} 
\end{figure}

The $\kappa(T)$ data measured along the $a$, $b$ 
and $c$-axes are shown in Fig.~\ref{K}.
Between 2 and 32~K, they are in qualitative agreement with unpublished 
results of Takeya \etal \cite{Takeya00thesis} but our $\kappa$-values 
are somewhat higher and the distinct features around $T_{N}=9.2\un{K}$, 
clearly visible in the insets of fig.~\ref{K}, 
are absent in the data of ref.~\cite{Takeya00thesis}. We suspect that 
this is due to the improved quality of the samples that were used in our study. 
%<<<<<<<<<<<<<<<<<<<<<<<< FIGURE 1 >>>>>>>>>>>>>>>>>>>>>>>>>
%<<<<<<<<<<<<<<<<<<<<<<<< FIGURE 1 >>>>>>>>>>>>>>>>>>>>>>>>>
%<<<<<<<<<<<<<<<<<<<<<<<< figure 1 >>>>>>>>>>>>>>>>>>>>>>>>>
Dips in $\kappa(T)$ upon magnetic ordering have often been observed  
in magnetic materials, and can be attributed to the scattering of 
phonons by enhanced fluctuations of the order parameter 
\cite{Kawasaki63,Stern65}.  
Below about 2~K, the thermal 
conductivity is almost isotropic  and the temperature dependence is $\kappa 
\propto T^{2.5}$. The anisotropy of $\kappa(T)$ 
is also weak  above about 170~K where, 
for all directions, $\kappa \propto T^{-0.5}$.
The anisotropy ratio $\kappa^{b}/\kappa^{a}$  for heat transport 
perpendicular to the chain direction is close to 1 and almost 
temperature-independent  across the entire covered temperature range. 
It seems reasonable to assume that perpendicular to the 
spin chain direction, the phonon heat transport dominates and
therefore, the weak anisotropy can easily be attributed to a weak 
anisotropy of the elastic constants and uncertainties in the evaluation of the 
sample geometry. 

At intermediate temperatures  the thermal conductivity $\kappa^{c}$
along the chain direction significantly exceeds those along the
perpendicular directions.
At 20~K, 
$\kappa^{c}/\kappa^{a}=2.3$ and $\kappa^{c}/\kappa^{b}=2.1$.  This 
particular and 
temperature-dependent anisotropy cannot be attributed to anisotropic 
phonon heat transport. 
Considering the layered crystal structure of BaCu$_2$Si$_2$O$_7$ with CuO$_4$ layers 
separated by Ba and SiO$_4$ layers, which are stacked along the  $b$-axis, 
it is expected that the phonon heat transport is weakest along the  
$b$-direction. Since this is not reflected in our observations, 
the anisotropy of the phonon heat transport 
is obviously very weak. A more detailed discussion of this point may be found in 
ref.~\cite{Sologubenko01}.  

\begin{figure}
\onefigure[width=0.65\columnwidth]{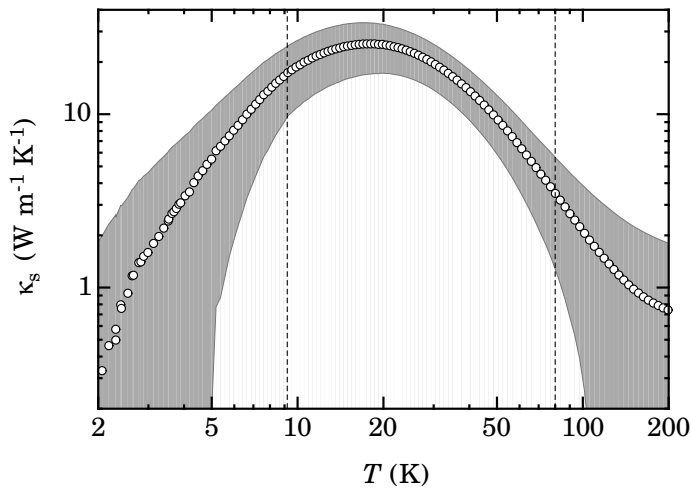}
\caption{Temperature dependence of the spin-mediated thermal 
conductivity  $\kappa_{s}(T)$ along the spin chains of 
BaCu$_2$Si$_2$O$_7$. The shaded area reflects the possible 
uncertainties which are caused by the subtraction of the phonon 
background contribution.}
\label{Ks} 
\end{figure}

The excess contribution to $\kappa$  along the $c$ direction is most 
naturally
associated with the 1D energy transport by itinerant spin excitations 
(spinons). 
In electrically insulating materials,  the total measured thermal 
conductivity along an $\alpha$-direction is 
$\kappa^{\alpha}(T) = \kappa_{ph}^{\alpha} + \kappa_{s}^{\alpha}$, where $\kappa_{ph}$ 
and  $\kappa_{s}$   are phonon and spin contributions, respectively. 
For directions perpendicular to the direction of weakly interacting 
spin chains, $\kappa^{a}_{s}, \kappa^{b}_{s} \approx 0$ because of the 
vanishingly small spinon  
velocities.  
To analyze the temperature dependence of
$\kappa_{s}^{c}$ (simply denoted as $\kappa_s$ below), 
we assumed that the phonon contribution $\kappa_{\rm 
ph}$ is nearly isotropic and subtracted $\kappa_{\rm ph} \approx 
(\kappa^{a} + \kappa^{b})/2$ from the total measured $\kappa^{c}(T)$.
The resulting values of $\kappa_s(T)$ are shown in fig.~\ref{Ks}. 
%<<<<<<<<<<<<<<<<<<<<<<<< FIGURE 2 >>>>>>>>>>>>>>>>>>>>>>>>>
%<<<<<<<<<<<<<<<<<<<<<<<< FIGURE 2 >>>>>>>>>>>>>>>>>>>>>>>>>
%<<<<<<<<<<<<<<<<<<<<<<<< figure 2 >>>>>>>>>>>>>>>>>>>>>>>>>
Since the measured values of $\kappa$ have maximum errors of 
about $\pm 10 \%$ along all directions, the related absolute maximum 
of possible uncertainties in  $\kappa_s(T)$, which are indicated 
by the shaded area in fig.~\ref{Ks}, are very large at high and low 
temperatures, where the  $\kappa^{\alpha}(T)$ curves for directions 
parallel and perpendicular to 
the chains differ only little. 
That is why we did not analyse the thermal conductivity in 
the AFM ordered state and 
restrict our analysis of $\kappa_{s}(T)$  to the limited 
temperature region between $T_{N}$ and 80~K (see dashed vertical lines in fig.~\ref{Ks}). 

Another reason for the observed anisotropy of $\kappa$ 
might be an anisotropy in the scattering strength between phonons and 
spinons. Although such a scenario cannot be completely ruled out, two 
facts suggest that it is very unlikely. 
Firstly, the nearly isotropic heat transport at the highest temperatures 
reached in this study is not compatible with a strong anisotropy of 
the phonon 
thermal conductivity and thus, the phonon-spinon interaction. 
Secondly, the observed reductions of the thermal conductivities at $T_{N}$, 
directly related to the spin-lattice interaction \cite{Kawasaki63}, 
are approximately the same 
for all crystallographic directions. Consequently, the features of $\kappa(T)$ at $T_{N}$, 
clearly seen in the insets of fig.~\ref{K}, are removed by subtracting 
$(\kappa^{a} + \kappa^{b})/2$ from $\kappa^{c}$. All this suggests 
that, although the scattering of phonons by spin excitations may be 
relatively effective, $\kappa_{\rm ph}$ 
remains, nonetheless, almost isotropic and hence the anisotropy of 
the measured $\kappa(T)$ is almost certainly due to heat transport by 
spin excitations along the $c$ direction. 

\begin{figure}
\onefigure[width=0.75\columnwidth]{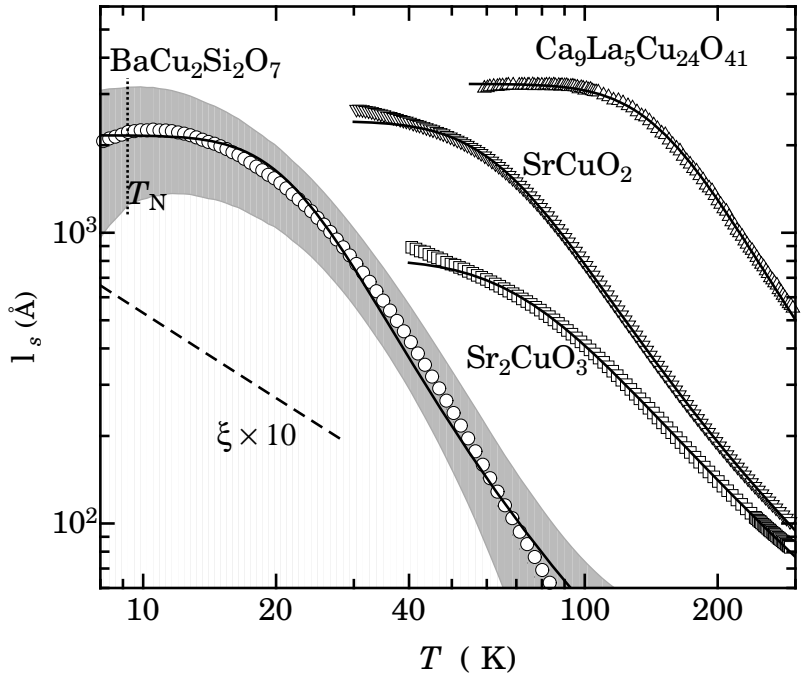}
\caption{Mean free path of spin excitations in BaCu$_2$Si$_2$O$_7$ in 
comparison with corresponding data for the spin-chain compounds Sr$_{2}$CuO$_{3}$ 
and  SrCuO$_{2}$ \cite{Sologubenko01}, and the two-leg spin-ladder material 
Ca$_{9}$La$_{5}$Cu$_{24}$O$_{41}$ 
\cite{Hess01}. The intrachain spin-spin correlation length $\xi$, 
shown as the broken line, is from 
ref.~\cite{Nomura91}}
\label{MagMFPs}
\end{figure}

From the values of $\kappa_{s(T)}$ and using eq.~(\ref{KsDrude}), we calculated the mean free path 
of the spin excitations $\ell_{s} = v_{s} \tau$, where $v_{s}(T)$ 
denotes the temperature-dependent  average velocity of propagating 
spin excitations. 
Assuming that the excitations of the Hamiltonian in
eq.~(\ref{Hamiltonian}) are spinons obeying 
Fermi-Dirac statistics
and adopting an energy dispersion \cite{Faddeev81}
 $ \varepsilon (k)= (J\pi / 2 ) {\sin kc}$,
where 
$c = 3.43\times 10^{-10}\un{m}$ 
is the distance between neighboring spins in a chain, 
$v_{s}(T)$ can be calculated as
\begin{equation}\label{Vs_ave}
  v_{s}(T)= {{1}\over{\hbar}} \left[ 
  \int {{\partial\varepsilon}\over{\partial k}}  f(\varepsilon,T) dk  \right] 
  / \left[ \int   f(\varepsilon,T) dk
  \right],
\end{equation}
where $f(\varepsilon,T) = (\exp (\varepsilon /k_B T)+1)^{-1}$.
The temperature dependence 
of the thermal Drude weight $D^{th}(T)$, taken from 
ref.~\cite{Kluemper02}, served to calculate $\tau(T)$.
The resulting values of $\ell_{s}(T)$ are shown in Fig.~\ref{MagMFPs}.
The shaded area again denotes the uncertainty. 
We note a rapidly increasing $\ell_{s}(T)$ with decreasing temperature 
and a trend to saturation at 
low temperatures. 

It is rather instructive to compare this $\ell_{s}(T)$ with available data for other 
similar insulating compounds.
Apart from BaCu$_2$Si$_2$O$_7$, mean free paths of spin excitations have 
been evaluated for the $S=1/2$ spin-chain compounds Sr$_{2}$CuO$_{3}$ and SrCuO$_{2}$ \cite{Sologubenko01}, 
CuGeO$_{3}$ \cite{Ando98,Takeya00,Salce98},  
and also for the undoped spin ladder compound Ca$_{9}$La$_{5}$Cu$_{24}$O$_{41}$. 
The data for Sr$_{2}$CuO$_{3}$ and  SrCuO$_{2}$ (data from \cite{Sologubenko01}), 
and  Ca$_{9}$La$_{5}$Cu$_{24}$O$_{41}$  \cite{Hess01} are shown in Fig.~\ref{MagMFPs}.
%<<<<<<<<<<<<<<<<<<<<<<<< FIGURE 3 >>>>>>>>>>>>>>>>>>>>>>>>>
%<<<<<<<<<<<<<<<<<<<<<<<< FIGURE 3 >>>>>>>>>>>>>>>>>>>>>>>>>
%<<<<<<<<<<<<<<<<<<<<<<<< figure 3 >>>>>>>>>>>>>>>>>>>>>>>>>
For CuGeO$_{3}$, the temperature dependence of $\ell_{s}$  was not 
calculated, but it was noted that $\ell_{s}$ does not 
exceed the value of $1300\un{\AA}$ \cite{Takeya00,Salce98}.
It is remarkable that
all $\ell_{s}(T)$ curves in fig.~\ref{MagMFPs} reveal a similar trend, 
in particular the low-temperature saturation 
value is always of the order of $10^{3}\un{\AA}$.
Also the high temperature features are similar and reveal a power law 
type $T^{-n}$ behavior with $n > 1$. 
It is important that, in spite of very different excitation spectra and 
therefore different temperature variations of $D^{th}(T)$ for spin chains and spin ladders 
\cite{Kluemper02,Alvarez02_Ano}, the corresponding extrinsic parameters 
$\ell_{s}(T)$ exhibit
similar features.  This suggests that the relaxation mechanisms 
for spin excitations are the same in all these compounds.

In view of the present lack of theoretical descriptions of
relaxation processes in spin chain 
compounds, only qualitative suggestions can be made. 
The first straightforward result is that the mean free path of spinons 
is not limited by the spin-spin correlation length. 
The intrachain correlation length in spin chains 
varies as $\xi \propto T^{-1}$ with small 
logarithmic corrections \cite{Nomura91}.
The result of calculations for BaCu$_2$Si$_2$O$_7$ 
is shown as the dashed line in Fig.~\ref{MagMFPs}. 
The absolute values of $\ell_{s}$ at low temperatures are considerably 
larger than $\xi$, and we
also note a much faster temperature-induced reduction of 
$\ell_{s}(T)$ than of $\xi(T)$ at elevated temperatures.

In ref.~\cite{Sologubenko00c}, a phenomenological approximation for 
the spinon mean free path was postulated as 
\begin{equation}\label{eSpinMFP}
\ell_s^{-1}(T)  = A_{sp} T \exp(-T^*/T) +  L_{sd}^{-1}.
\end{equation}
The first term on the right-hand side reflects the scattering of 
spinons  by phonons, with the  parameter  $A_{sp}$   
characterizing the  spin-lattice interaction strength and
$T^*$ representing a threshold energy typical for Umklapp-processes.
The second and constant term is 
associated with the influence of magnetic defects due to non $S=1/2$ 
ions on the Cu sites, with  
$L_{sd}$ as the mean distance between them.
Eq.~(\ref{eSpinMFP}) is equally well applicable  
to all data sets shown in Fig.~\ref{MagMFPs}. 
The solid lines in Fig.~\ref{MagMFPs} are fits to 
eq.~(\ref{eSpinMFP}) with the parameter values given in 
Table~\ref{Table1}.
%========= Table ============
\begin{table}
\caption{Parameters of the fitting of $\ell_{s}(T)$ data to 
eq.~(\ref{eSpinMFP})  }
\label{Table1}
\begin{center}
\begin{tabular}{lcccc}
Parameter  & Sr$_{2}$CuO$_{3}$ & SrCuO$_{2}$  & Ca$_{9}$La$_{5}$Cu$_{24}$O$_{41}$  
& BaCu$_2$Si$_2$O$_7$    \\
$L_{sd}$ (nm)                & $81\pm 2$  & $242\pm 2$  & $325\pm 
1$ &  $216\pm 2$ \\
$A_{sp}$ ($10^{5}\un{~m^{-1}}$) & $7.1\pm 2$ & $6.7\pm 1$  & $3.3\pm 
0.2$ & $41.4\pm 2$ \\
$T^*$ (K)                    & $177\pm 5$ & $204\pm 2$  & $532\pm 10$ & 
$80\pm 2$ 
\end{tabular}
\end{center}
\end{table}
%---------- Table -----------
If $L_{sd}$ indeed reflects the limitation of the mean free path 
by defects, the  fit values of this parameter 
given in Table~\ref{Table1} 
are the consequence of similar densities of these defects in these
materials. 
The exception is  Sr$_{2}$CuO$_{3}$ where the smaller value 
of $L_{sd}$ might be due to the additional influence of 
bond-defects \cite{Sologubenko00c}.

The relatively large value of the parameter $A_{sp}$ for BaCu$_2$Si$_2$O$_7$ 
indicates enhanced spinon-phonon scattering in this compound. This can be 
understood by taking into account that for this material, the intrachain 
exchange constant $J/k_{B} = 279\un{K}$ is close to  the value of 
the Debye temperature $\theta_{D}$ which, from 
low-temperature specific heat data presented in ref.~\cite{Yamada01}, 
is approximately 290~K. The near identity of the energy scales for spin and 
lattice excitations leads to stronger scattering processes 
between the two types of quasiparticles. For the other compounds, the values of 
$J/k_{B}$ are about an order of magnitude large than $\theta_{D}$. 

If the strongly increasing $\ell_{s}(T)$ with decreasing $T$ is indeed 
due to the reduction of spinon-phonon scattering, it may happen that at low 
enough temperatures the heat input into a sample will produce a 
temperature gradient in the crystal lattice only. Correspondingly, the spin system 
adopts a constant average temperature and does not participate in the 
energy transport processes \cite{Sanders77}.  In ref.~\cite{Sologubenko01} 
we estimated that for Sr$_{2}$CuO$_{3}$ and SrCuO$_{2}$ this happens 
quite abruptly in the temperature region between about 10 and 20~K. However, the 
wavevector dependence of the spinon-phonon interaction was not taken into 
account. Such a dependence would lead to a broadening of the transition 
from the high-temperature region, where $\kappa_{s}$ can be measured 
by our methods, to the low-temperature regime, where $\kappa_{s}$ in 
not observable in this way. In principle, the low-temperature saturation of 
$\ell_{s}(T)$ may reflect such a transition, instead of the 
influence of defects.

Although eq.~(\ref{eSpinMFP}) captures the main features of the temperature 
variation $\ell_{s}(T)$, it is not properly justified theoretically. A 
rigorous 
quantum mechanical analysis of the influence of the spin-lattice 
interaction on the energy relaxation in quantum spin chains would definitely 
be of great help.

In conclusion, the analysis of the experimental results of anisotropic heat 
transport in BaCu$_2$Si$_2$O$_7$ allowed for the evaluation of the mean free path $\ell_{s}(T)$
of itinerant spin excitations. A comparison with several other compounds 
containing isotropic Heisenberg  $S=1/2$ chain-type structural 
elements demonstrates 
that $\ell_{s}(T)$ of all these materials exhibit similar features, 
and indicates that a universal type of energy relaxation for spinons 
is dominant in these materials.  

\acknowledgments
This work was financially supported in part by
the Schweizerische Nationalfonds zur F\"{o}rderung der Wissenschaftlichen
Forschung.

%\bibliography{1227,AddRefs}

\end{document}